\documentclass[conference]{IEEEtran}

\usepackage{cite}

\usepackage{graphicx}
\usepackage{amssymb,amsmath,bm}
\usepackage{textcomp}
\usepackage{float}
\usepackage{placeins}
\usepackage{color}
\usepackage{multirow}
\usepackage{booktabs}

\def\vec#1{\ensuremath{\bm{{#1}}}}

\title{Acoustic Modeling Using a Shallow CNN-HTSVM Architecture}

\author{\IEEEauthorblockN{Christopher Dane Shulby$^1$$^,$$^2$, Martha Dais Ferreira$^1$, Rodrigo F. de Mello$^1$, Sandra Maria Aluisio$^1$}
\IEEEauthorblockA{
  $^1$Institute of Mathematical and Computer Sciences \\ Univeristy of Sao Paulo, Brazil\\
  $^2$CPqD, Brazil \\
  Email: {\tt 
cshulby@icmc.usp.br, daismf@icmc.usp.br, mello@icmc.usp.br, sandra@icmc.usp.br}}} 
\makeatletter
\def\name#1{\gdef\@name{#1\\}}
\name{{\em Christopher Dane Shulby$^1$$^,$$^2$, Martha Dais Ferreira$^1$, Rodrigo F. de Mello$^1$, Sandra Maria Aluisio$^1$}}

\begin{document}

  \maketitle
  \begin{abstract}
High-accuracy speech recognition is especially challenging when large datasets are not available. It is possible to bridge this gap with careful and knowledge-driven parsing combined with the biologically inspired CNN and the learning guarantees of the Vapnik Chervonenkis (VC) theory. This work presents a Shallow-CNN-HTSVM (Hierarchical Tree Support Vector Machine classifier) architecture which uses a predefined knowledge-based set of rules with statistical machine learning techniques. Here we show that gross errors present even in state-of-the-art systems can be avoided and that an accurate acoustic model can be built in a hierarchical fashion. The CNN-HTSVM acoustic model outperforms traditional GMM-HMM models and the HTSVM structure outperforms a MLP multi-class classifier. More importantly we isolate the performance of the acoustic model and provide results on both the frame and phoneme level considering the true robustness of the model. We show that even with a small amount of data accurate and robust recognition rates can be obtained.
 
  \end{abstract}

  \section{Introduction}
  
    Most speech processing applications rely on acoustic models which build the bridge between the audio signal and its phonetic transcription. %After a sentence prompt has been phonetically transcribed, the task of learning which phones go with which audio segments is far from trivial and essential to modern speech applications since problems at this stage are likely to propagate even with the help of a robust language model. For automatic speech recognition (ASR) it is essential that phonemes are transcribed completely and in the correct order, in other words all phonemes in sequence should be recognized. Normally an ASR system will consist of an acoustic model, a pronunciation model and a language model where the acoustic model tries to match the signal to its probable phoneme(s) and these posterior values are used in the pronunciation model to find the most likely words for each segment which is then sent to the language model to determine the most likely chunks or phrases.
After a sentence prompt has been phonetically transcribed, the task of learning which phonemes belong to certain audio segments is far from trivial and essential to modern speech applications, since problems at this stage are likely to propagate even with the help of a robust language model. For Automatic Speech Recognition (ASR), it is essential that phonemes are transcribed completely and in the correct order, in other words, all phonemes in sequence should be recognized. Normally, an ASR system consists of an acoustic model, a pronunciation model and a language model. The acoustic model attempts to match the signal to its probable phoneme(s) and the posterior values are used in the pronunciation model to find the most likely words for each segment, which works as input to the language model to determine the probable chunks or phrases.

    Most studies report the accuracy of systems after using all these components; however, it is important to accurately model the acoustic properties so that errors do not propagate in the other models. Reliable phoneme-level error detection is still a great need~\cite{witt2012automatic,li2017mispronunciation} for automatic pronunciation training where a pronunciation model can actually damage the desired results.
    
    The long time state-of-the-art GMM-HMM acoustic models have become less popular since the Deep Learning (DL) movement in the last five years or so. In many cases, DL has significantly reduced the error rates of ASR  systems~\cite{hinton2012deep}. Still, the complexity of neural network models can be a problem, since they must be trained with huge corpora. Recent corpora released for DNN (Deep Neural Network) models contain over a thousand hours of audio data~\cite{cieri2004fisher, panayotov2015librispeech} and can take months to train~\cite{chen2012pipelined}. GMM-HMM models have the advantage that ``good'' models can be built using relatively little data (in few hours) and they do not require the enormous corpora needed for training DNNs~\cite{chan2005sphinx}. The trade-off in quality vs. quantity seems to be somewhere on the spectrum between these two methods for most applications. Besides the training time, the real problem lies in the availability of training data for under-resourced languages or specific applications.
    
    %In this paper we propose a method which could be useful for low-resource training environments. In this paper we present a Shallow-CNN-HTSVM (Hierarchical Tree Support Vector Machine classifier) architecture which combines several techniques which have had success in ASR tasks.
In this paper we propose a method useful for low-resource training environments, in which a Shallow-CNN-HTSVM 
(Hierarchical Tree Support Vector Machine classifier) architecture is used. This architecture combines several techniques which have already shown success in ASR tasks.   
We have done this by taking a knowledge-driven slant on the typical machine learning (ML) approach. First, we take advantage of the Convolutional Neural Networks (CNNs) well-known ability to deal with images~\cite{lecun1998gradient, krizhevsky2012imagenet} which extract the features from a spectrogram generated from the audio signal. The selection of the CNN is in the spirit of a knowledge-driven manual classification task where trained specialists in acoustics and laboratory phonetics~\cite{ladefoged2012vowels} are able to classify phonemes even when the spectrogram is fairly noisy. This is due to the visual representation given to us via Fast Fourier transform.
%(FFT)
Humans are able to see a ``picture'' based on the relative heights and intensities of formants and energy concentrations across spectrogram frequencies. While the actual frequency values can vary greatly from speaker to speaker depending on multiple factors including the length of the vocal tract, the ``picture'' is always similar and recognizable to a human scientist. Therefore, it makes sense that a computer vision algorithm would be suitable in this case.
    %The down sampling of a CNN is actually inspired  by the  cat’s  primary  visual  cortex \cite{hubel1962receptive},  which  identified orientation selective simple cells with overlapping local receptive fields, or sub regions. It operates locally \cite{abdel2014convolutional} convolving with filters over and image where it is able to recognize similar avatars, regardless of their actual position. We believe that the use of filter masks could be another useful feature which the CNN offers to phoneme recognition. These are some advantages over other neural networks which are unable to deal with this type of translational variance local distortions of the input, as is pointed out in \cite{lecun1995convolutional}. This gives the CNN a higher level of robustness against distortions due to speaker variability and noise, both of which are much needed in the current state-of-the-art ASR applications.
    
	We classify the features using a hierarchical tree with predefined articulatory groups where each node contains an SVM (Support Vector Machine). The SVM was selected because it provides supervised learning guarantees due to the VC theory~\cite{vapnik1998statistical} and the principle of structural risk minimization. The hierarchical tree structure was chosen due to the unbalanced data problem where we believe that a good way to work with this problem is to combine a knowledge-driven recipe with a ML algorithm, thus simplifying the classification space while saving time, since SVMs with a large amount of samples can become prohibitively costly to train. The goal of this study is to present an acoustic model with high accuracy given limited resources, in both senses of computational processing power as well as training data. 
  
\section{Related Work}

	Phoneme recognition is not a new task as explored in~\cite{waibel1989phoneme, lee2009unsupervised} and~\cite{hau2011exploring}, but greater success has been achieved only in the last five years or so~\cite{hinton2012deep} and still remains far from a solved problem. One of best known studies proving the capabilities of CNNs for this task is~\cite{abdel2012applying}, where a hybrid CNN-HMM model using local filtering and max-pooling in the frequency domain is proposed to deal with the translational invariance problem present in other DNNs. In~\cite{sainath2013deep}, the optimal CNN architecture is explored including the number of convolutional layers and hidden units needed, as well as the optimal pooling strategy and feature type for CNNs and best results are achieved using large corpora (300-400 hours) and a two-convolutional-layer DNN with with 424 hidden units and four fully connected layers with 2,048 hidden units each, followed by a softmax layer with 512 output targets.
    
    Abdel-Hamid, et al.~\cite{abdel2014convolutional} revisits the issue of robustness in speaker and environment variation with a CNN-HMM where the HMM deals with the issue of distortions of over time, while the CNN convolves over the frequency to take advantage of its ability to deal with variations among speakers in this domain. This study serves as a baseline on TIMIT for the state-of-the-art deep CNN with a 21.6\% PER (Phone Error Rate).
Here the authors do not explain how the strings are generated for comparison with the original TIMIT annotations but one can assume that some post-processing is done to reduce the frames to phones. A general issue, for proper comparison of acoustic models, is that state-of-the-art methods which use large deep networks with thousands of units and often thousands of hours of training data, do not show frame-level results and as in~\cite{mohamed2012acoustic,abdel2012applying,sainath2013deep,abdel2014convolutional,graves2014towards,hannun2014first,toth2015phone}. As outlined in \cite{hinton2012deep}, they often employ a number of resources like pronunciation models, language models and other post-processing/data smoothing techniques which are of great help for the end speech-recognition applications; however, they also mask the true recognition accuracy achieved by the acoustic model.

Some studies do report FER (Frame Error Rate) which represents the true accuracy of the acoustic model. In~\cite{lombart2014articulatory} a hybrid CNN and layer-fused MLP (Multi-layered Perceptron) with inputs of 11 frames of 25ms (step of 10ms) as context from the TIMIT database and presents both PER and FER (Frame Error Rate) where the best network with 1024 neurons in the input layer and 512 in the hidden layer and were able to obtain a FER of 43.04\%. In \cite{lopes2009phonetic}, a hierarchical broad-phoneme MLP/HMM hybrid classifier was used with window widening which achieved impressive results. For a 90ms window the FER was 61\% using their past-future method that achieved 42\% using 170ms and 17 frames for context training every other frame along the right and left side of the central frame for that period. It should be noted that these experiments were done with the full 61 phoneme set and results for smaller windows with 39 phonemes are not presented.

    Hierarchical classification has not been often applied to the phoneme recognition task but some notable exceptions exist, like~\cite{dekel2004online, karpagavalli2015hierarchical, driaunys2015implementation} and \cite{amami2015study}. In~\cite{dekel2004online}, phoneme classification is treated as an optimization problem where a hierarchical tree structure divides groups of phonemes as nodes on the tree. The authors found that the tree would tolerate small tree-induced errors while avoiding gross errors as a standard multi-class classifier would be prone to commit. In the last two years, the HTSVM has been employed using data from speech corpora and applied to a phoneme recognition task as presented in~\cite{driaunys2015implementation}, \cite{karpagavalli2015hierarchical}, and \cite{amami2015study}. In~\cite{driaunys2015implementation}, an experiment on stop and fricative consonants using the Lithuanian LTDIGITS corpus containing over 25,000 phonemes. The most important findings were a 3\% gain in the overall accuracy, a total of 68.4\%, while reducing 52-55\% of the computational time taken for classification with SVMs. In~\cite{karpagavalli2015hierarchical}, a tamil corpus of repeated words was developed and 2,400 phoneme instances were tested resulting in about 67\% total accuracy on obstruent and sonorant sounds using MFCC features. A study on the TIMIT corpus~\cite{amami2015study}, presents MFCC classification results for each phoneme and major confusions classified by SVMs as well. Most of the phonemes fall between the accuracy range of 30\% and 60\%. The authors point out that due to the multiple dialects present in the TIMIT corpus, many phonemes are pronounced similarly to others depending on the speaker, increasing the confusion rate for similar phonemes where the SVM was not capable of efficiently classifying phonemes in the lower nodes of the tree.

\section{Dataset, Features and ML Algorithms}
     
\subsection{Dataset: TIMIT automatic segmentation}
     
%\subsubsection{TIMIT}
     
The TIMIT Corpus
%developed by Texas Instruments (TI) and the Massachusetts Institute of Technology (MIT)~\cite{garofolo1990darpa}, 
is a speech corpus meant to be used for speech research and to serve as a standard for results comparison. TIMIT contains phonetically balanced prompted recordings of 2,342 unique sentences (2 dialect sentences (SA), 450 phonetically compact sentences (SX) and 1,890 phonetically-diverse sentences (SI)) from 630 speakers of eight major dialects of American English, each reading ten phonetically rich sentences to total 6,300 utterances (5.4 hours). The full training set contains 4,620 utterances. The test set contains 1,344 utterances from 168 speakers. With the exception of SA sentences which are usually excluded from tests, the training and test sets do not overlap and follow the suggested corpora splits outlined in~\cite{garofolo1990darpa}.
TIMIT is considered a ``balanced'' corpus with respect to the distribution of phones and triphones found in the English language but it also follows a typical Zipf curve as one would expect to find in a speech corpus, where some phoneme sequences dominate a large portion of the samples and a large number of phonemes are less frequent. From a machine learning perspective, this is an unbalanced dataset. Normally, as we have done here, the phone set is collapsed into 39 monophones as suggested in~\cite{lee1989speaker}.

%\subsubsection{TIMIT automatic segmentation}

	In order to build a full pipeline for low-resourced language situations the label generation was delegated to an automatic labeler, pretrained with only one hour of data using HTK~\cite{young2002htk}. In order to improve the performance of this tool, some adjustments were made, mainly a rule based script was created to generate multiple pronunciations for each word in the pronunciation dictionary in order to account for co-articulation and the eight dialects used in the TIMIT database and some light manual revision of difficult cases was done. Then the labeler was trained on the rest of the TIMIT training set.

\subsection{Spectrogram Images}
     
\begin{figure}[h]
\centering
\includegraphics[scale=0.6]{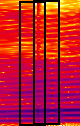}
\caption{{\it Spectrogram illustrating 25ms Hann windows with a stride of 10ms.}}
\label{fig:stride}
\end{figure}

	For each audio file, spectrograms of $25$ms Hann windows with a stride of $10$ms were extracted, as shown in Figure~\ref{fig:stride}. The images were resized to $5 \times 128$ pixel images. This was done in order to reduce the number of features extracted to a manageable dimensionality while prioritizing the frequency information. When a sentence ended in a number of milliseconds which is not divisible by $25$, the last frame was squeezed into the penultimate frame (in all cases, this was silence). This process yielded $1,447,869$ images for training and $482,623$ for testing.

\subsection{Feature Extraction}

CNN is a deep learning technique, which presents good results in several domains~\cite{lecun1995convolutional}, including speech processing~\cite{abdel2012applying, sainath2013deep,abdel2014convolutional}. As mentioned earlier, the biological motivations prompted us to employ a CNN to extract relevant features from the spectrogram images, conserving only the most important information. The advantage of the CNN in this application is the identification of local recurrence information and its invariant translation in data~\cite{lecun1998gradient, lecun15, Goodfellow16}. CNNs are tolerant to distortion as they combine local receptive fields, shared weights and spatial sub-sampling. All three of which are useful in phoneme recognition~\cite{abdel2012applying}. The trick comes in balancing the spatial resolution reduction with the representational richness of the images in order to generate the most useful feature maps at a low classification cost with high accuracy~\cite{lecun1998gradient}. We perform this in two ways: first, by rescaling the image to a smaller size while still preserving the most important information for phoneme classification and, secondly, by searching for the best sized masks. It should be noted that we use only single frames for classification.

The shallow CNN architecture used in our experiments was empirically set after several experiments, considering just one convolutional and sub-sampling layer. The best configuration found has $38$ convolutional units with a $29 \times 1$ mask size, and $38$ units of max-pooling in the sub-sampling operation sized $5 \times 5$ without overlapping. ReLU (Rectified Linear Unit) was applied as the activation function due to its widespread adoption in literature. %This function avoids negative values and maintains the scale of output values. 
It took less than half an hour to execute the CNN on a computer with 8GB of RAM and conventional hardware (Intel i7).
%Figure~\ref{fig:cnn} illustrate the CNN architecture.

%\begin{figure}[H]
%\centering
%\includegraphics[width=0.5\textwidth]{cnn.eps}
%\caption{{\it CNN architecture defined for the experiments.}}
%\label{fig:cnn}
% * <chrisshulby@gmail.com> 2017-02-28T18:25:18.619Z:
%
% ^.
%\end{figure}

\subsection{Classification}

The proposed method takes advantage of the machine learning techniques used for CNNs and SVMs and attempts to improve accuracy and minimize their disadvantages with a knowledge-based hierarchical tree structure. The features produced by the CNN were classified using a SVM. 
%SVM has been used in literature combined with CNN that has been provided good results in many domains~\cite{}. In addition, 
SVM provides a strong learning guarantee according to the Statistical Learning Theory and large-margin bounds~\cite{vapnik1998statistical, von2008statistical}. The SVM parameters were found empirically after several experiments. The selected kernel for final experiments was a $4$\textsuperscript{th} order polynomial kernel with $coef0=1$ (as a non-homogeneous kernel) and a cost $C=10,000$. For this task, namely speech recognition on the TIMIT dataset, we have two main problems to solve. First, the number of CNN extracted features combined with the number sample frames, since the SVM training time increases quadratically as the number of examples increases. The second problem is the unbalanced nature of the dataset, a common characteristic for most speech corpora.

\begin{figure}[H]
\centering
\includegraphics[width=0.4\textwidth]{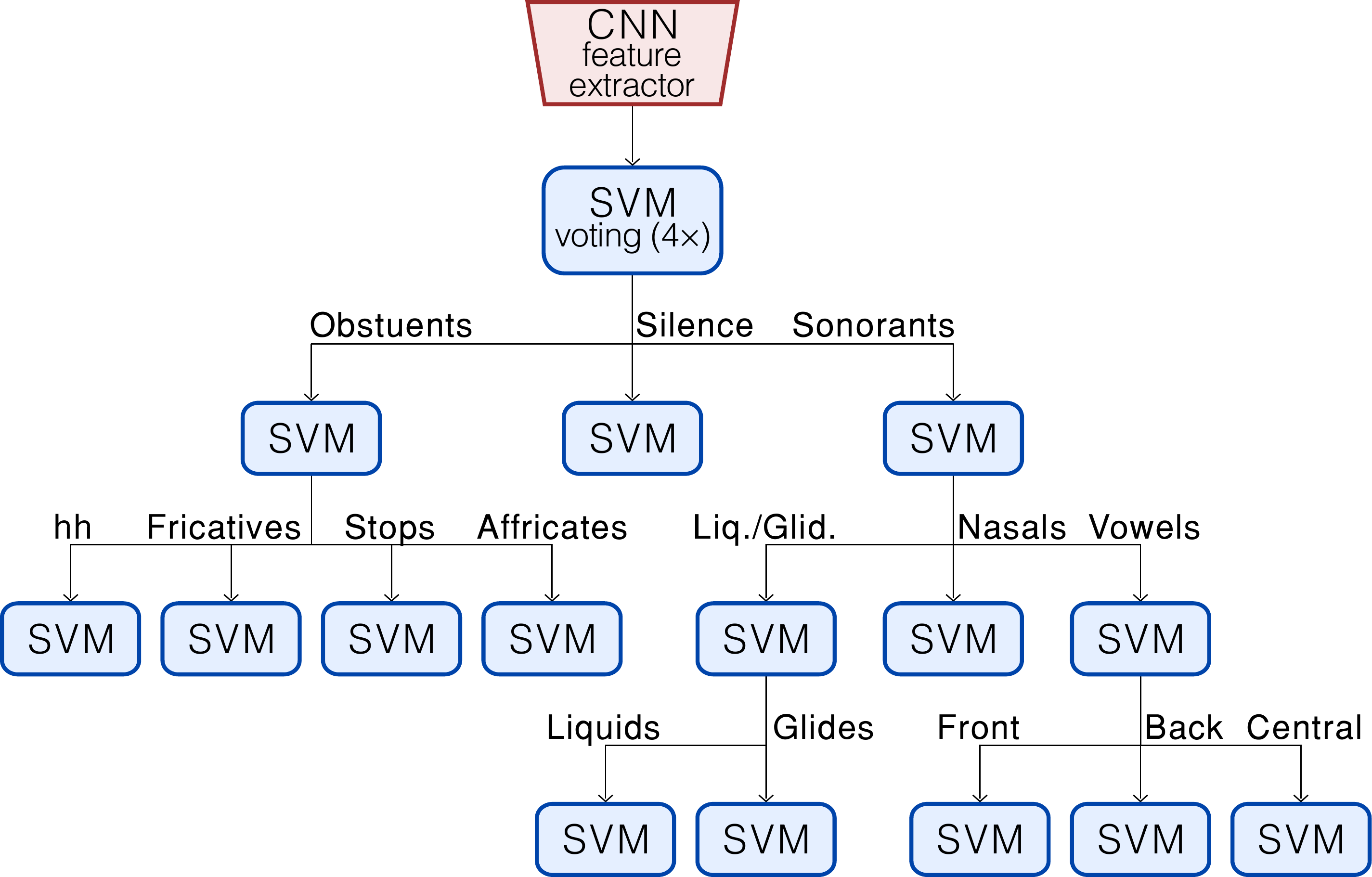}
\caption{{\it HTSVM architecture defined for the experiments.}}
\label{fig:htsvm}
\end{figure}

     For the first issue, the hierarchical structure is what makes the SVM a viable option. The cost of training a sequential SVM on this dataset of almost 1.5 million images with our 988 dimensions would be prohibitive and even with a great deal of work in data reduction techniques, it would still likely take several months to train the model. By dividing the task into several hierarchical levels based on the knowledge of articulatory phonetic classifications in English as described in~\cite{ladefoged2012vowels}, we are able to turn the problem into a binary, ternary or quaternary classification instead of the original 40 classes. Ladefoged suggests a hierarchical structure necessary for English features in the last chapter of~\cite{ladefoged2012vowels}. This served as the primary source for our tree which was derived as possible questions to classify each phoneme so that they contain the features necessary for classification. This makes our classification space much more simple when creating the support vectors. It should be noted that the first layer classifying obstruents, silence and sonorants is built using 5 individually trained SVMs on equal chunks of data where the prediction for this layer is made by a simple voting system where the mode is taken as the final prediction. This was done to further reduce the training time. The hierarchical structure is presented in Figure~\ref{fig:htsvm}.
      
     In the second issue, we deal with an unbalanced dataset where even minimal pairs can have a large difference in frequency as in the example of /k/ and /g/ which are both velar stops. In the training set, the phoneme /k/ appears in $60,433$ frames, whereas /g/ is found in only $17,727$. In order to build a robust system, it is important to learn this phonemic distinction and minimize the influence of probability in the training set. We were able to deal with this by using the SMOTE~\cite{chawla2002smote} data augmentation technique. The synthetic creation of minority samples allows us to treat the classification task with more confidence. This was also made possible by the hierarchical approach because each node has a much smaller number of samples than the original dataset.

\section{Experiments}
     
     	As a baseline comparison to the proposed method, we used one of the most popular ASR toolkits for a database the size of TIMIT, the HTK toolkit~\cite{young2002htk}. A triphone HTK model with $31$ Guassians was trained on the same TIMIT training set used in our method and recognition was performed on the test set with a zero-gram language model and only the individual monophones as the pronunciation model. This was done in order to obtain only the posterior values from the acoustic model predictions without the influence of a language or pronunciation model for fair comparison, since we are only evaluating the accuracy of the acoustic model. The predictions were then segmented in the same fashion as the proposed method with $25$ms sliding windows and a step of $10$ms, so that a frame by frame comparison could be made. Along with the HTK model, we also trained a simple MLP with one hidden layer and $100$ neurons and a ReLu activation function. This was done to show the gain provided by the HTSVM structure over another widely used classifier for the classification of CNN features.
        
        For each model, we calculated the FER and F1 score since accuracy can be misleading at times when dealing with a number of less frequent phonemes. This was done first on a frame by frame analysis. Then we smoothed the data by removing frame repetitions in order to collapse them into sequential phonemes. For the later, we were also able to calculate the PER (Phoneme Error Rate), the industry standard for measuring the accuracy of acoustic models and is calculated by the the Levenshtein distance~\cite{levelshtein-66-binary} where the number of insertions, deletions and substitutions were added and divided by the total number of phonetic units in the string. It should be noted that for fair comparison, we use the same 39 phonemes used in other studies, omitting silence.
        
        Table~\ref{tab:acc_f1} presents the F1 scores and FER of the GMM-HMM, CNN-MLP and CNN-HTSVM models for frame classifications as well as the PER in the case of phoneme classification.
        
              \begin{table}[H]
        \centering
        \caption{\it F1 Scores in Frames, Frame Error Rates and Phone Error Rates for each model.}
        \label{tab:acc_f1}
        \footnotesize
        \center{
          \begin{tabular}{ ccccccc }
           % \hline
            \toprule
            \multicolumn{1}{c}{Classifier} & &
            \multicolumn{1}{c}{F1 Score} & &
            \multicolumn{1}{c}{FER\%} & &
            \multicolumn{1}{c}{PER\%} \\
            \cmidrule{1-1} \cmidrule{3-3} \cmidrule{5-5} \cmidrule{7-7}
              GMM-HMM & & 0.166 & & 76.36 & & 75.17 \\
              CNN-MLP & & 0.225 & & 56.97 & & 52.90 \\
              CNN-HTSVM & & 0.491 & & 37.04 & & 35.41 \\
            \bottomrule
          \end{tabular}
        }
      \end{table}
              
      Independent of the models accuracy, it is also important to understand what sort of errors the models are actually committing. Due to space limitations, Table~\ref{tab:phone_conf_1} 
%      and \ref{tab:phone_conf_2} 
only lists the $10$ most frequent errors committed by each system, including the true values, predicted values and the confusion percentage.  
       
       \begin{table}[!htbp]
       \centering
       %\parbox{.85\linewidth}{
        \caption{\it Most Frequent FER Confusion percentages in GMM-HMM and CNN-HTSVM models where the true phoneme was confused as being the predicted phoneme.}
        \label{tab:phone_conf_1}
        \footnotesize
        \center{
          \begin{tabular}{ ccccccc }
            \toprule
            \multicolumn{3}{c}{GMM-HMM} & & \multicolumn{3}{c}{CNN-HTSVM} \\
            \cmidrule{1-3}  \cmidrule{5-7}
              True & Pred &  Conf (\%) & & True & Pred &  Conf (\%) \\
              \midrule
              s & z &  33.14 & & s & z & 15.16 \\
              ih & uw & 16.00 & & ay & ae & 39.64\\
              t & ch & 17.58 & & ao & aa & 26.58\\
              %k & sil & 25.26\\
              er & r & 32.46 & & r & er & 18.84 \\
              %t & sil & 22.06\\
              ao & l & 28.00 & & sh & s & 26.01 \\
              iy & y & 14.23 & & aa & ae & 16.07 \\
              %p & sil & 32.73\\
              s & sh & 10.09 & & ah & ae & 14.79\\
              ae & t & 14.32 & & t & s & 7.61 \\
              ih & z & 10.07 & & iy & ih & 6.48\\
              w & ao & 45.52 & & er & r & 7.64 \\
              %sil & t & 3.78 \\
         %     iy & uw & 12.80 & & er & r & 10.64 \\
              %sil & k & 3.76\\
         %     k & eh & 11.55 & & z & s & 18.01\\
          %    ih & t & 7.70 & & ay & aa & 15.78 \\
              %sil & s & 3.40\\
           %   ah & l & 16.67 & & t & s & 10.61 \\
            %  d & t & 14.46 & & iy & ih & 9.48\\
            %  d & uw & 14.16 & & er & r & 10.64 \\
             %sil & sh & 4.14 & & z & s & 18.01\\
             % d & uw & 14.16 & & ay & aa & 15.78 \\
            % ih & ng & 6.96 & & ih & iy & 6.33\\
           %  iy & z & 8.94 & & er & eh & 8.06 \\
            \bottomrule
          \end{tabular}
        }
%        }
      \end{table}
%        \begin{table}[!htp]
%        \parbox{.85\linewidth}{
%         \caption{CNN-HTSVM Most Frequent Confusions.}
%         \label{tab:phone_conf_2}
%        \footnotesize
%         \center{
%           \begin{tabular}{ |c|c|c| }
%             \hline
%             \multicolumn{1}{|c}{True Phoneme} &
%             \multicolumn{1}{|c|}{Predicted Phoneme} & 
%             \multicolumn{1}{c|}{Confusion \%} \\
%             \hline \hline
%               s & $z $&       $18.16\%$~~~ \\
%               ay & $ae $&       $42.64\%$~~~ \\
%               k & $sil $&       $25.26\%$~~~ \\
%               t & $sil $&       $22.06\%$~~~ \\
%               ao & $aa $&       $29.58\%$~~~ \\
%               p & $sil $&       $32.73\%$~~~ \\
%               r & $er $&       $21.84\%$~~~ \\
%               sh & $s $&       $29.01\%$~~~ \\
%               aa & $ae $&       $19.07\%$~~~ \\
%               sil & $t $&       $3.78\%$~~~ \\
%               sil & $k $&       $3.76\%$~~~ \\
%               ah & $ae $&       $17.79\%$~~~ \\
%               sil & $s $&       $3.40\%$~~~ \\
%               t & $s $&       $10.61\%$~~~ \\
%               iy & $ih $&       $9.48\%$~~~ \\
%               er & $r $&       $10.64\%$~~~ \\
%               z & $s $&       $18.01\%$~~~ \\
%               ay & $aa $&       $15.78\%$~~~ \\
%               ih & $iy $&       $6.33\%$~~~ \\
%               er & $eh $&       $8.06\%$~~~ \\
%             \hline
%           \end{tabular}
%         }
%         }
%  %    \clearpage 
     
     The confusion percentage for each true class is defined as the percentage of all occurrences which were misclassified as predicted class.

\section{Discussion}
  
Most errors committed by the HTSVM were made between similar phonemes as found in~\cite{amami2015study}, many of which unlikely would be perceived by human listeners. For example, the confusion between /s/ and /z/ are likely produced by the database itself where speakers may mix the two or use them interchangeably where it is semantically unimportant~\cite{hoffmann2009automatic}. This shows that the algorithm is quite promising since it avoided gross errors which are common in many modern acoustic modeling algorithms like GMM-HMM. The issue between similar phonemes seems to be one which could be corrected by a robust language and/or pronunciation model. Also, since such a small window of $25$ms was used, some of the vowels, especially diphthongs like /ay/ were damaged. Studies like~\cite{karpagavalli2015hierarchical} use $9$ frames for classification and \cite{yang2000relevance} suggest at least $100$ms for context which prompted~\cite{lombart2014articulatory} to use $11$ frames ($110$ms) as context for classification. While our system is more accurate in frame classification than the system in~\cite{lombart2014articulatory}, our PER is worse. This is probably also due to the small size where unnecessary insertions are made in the phoneme strings.

	 The combination of features extracted from single frames by a simple and shallow CNN classified by the HTSVM produced similar accuracy results to that study on a more difficult task as we were classifying full utterances which include pauses, co-articulation and a greater variation in pronunciation and not only single words and likely more careful speech. It is also important to note that some errors, like the case of the vowels which are close in the vowel space, could have been provoked by the TIMIT transcriptions themselves where in some dialects these sounds could be very similar or even some sound in between.

\section{Conclusions and Future Work}

We have shown that even with the large dimensionality of the CNN features, a shallow CNN-HTSVM architecture can be useful in scenarios where a large abundance of data is not available. Our method shows similar or better results than the most recent HTSVM methods and outperforms other works using few frames for classification as well as the traditional GMM-HMM classifier and an MLP classifier using the same features. We also make a contribution towards the comparison of acoustic models by presenting a breakdown of frame and phoneme accuracy with F1 scores which were not available in many of the previous studies, providing more information about the true robustness, facilitating the assessment of state-of-the-art acoustic models.

As future work, our CNN-HTSVM pipeline could take advantage of some fine-tuning. Since most errors were made between very similar phonemes and diphthongs, it would be interesting to investigate this issue further. As a first step, one could implement something along the lines of a backtracking system in the tree where after a preliminary decision is made in the node (especially in the final node), a second node could check for likely errors according to predefined rules. As a second solution to this problem, one could add more frames for the classification stage as employed in other studies. It also would be useful to investigate other features which could be added to better disambiguate some of the more frequent errors, for example, acoustic features to indicate voicing for sonorants and some fricatives or voice onset time between stops and fricatives. We also believe that the CNN will be even more valuable when applied to a noisy database or a database with more classes and we are currently exploring this scenario.

%\section{Acknowledgements}
  
%We would like to thank Dr. Rodrigo Mello's mother for teaching him machine learning and his grandmother for teaching him liner algebra, both of which he then passed down to us.

 % \newpage

  
\bibliographystyle{IEEEtran}

  \bibliography{mybib}

%  \begin{thebibliography}{9}
%    \bibitem[1]{Davis80-COP}
%      S.\ B.\ Davis and P.\ Mermelstein,
%      ``Comparison of parametric representation for monosyllabic word recognition in continuously spoken sentences,''
%      \textit{IEEE Transactions on Acoustics, Speech and Signal Processing}, vol.~28, no.~4, pp.~357--366, 1980.
%    \bibitem[2]{Rabiner89-ATO}
%      L.\ R.\ Rabiner,
%      ``A tutorial on hidden Markov models and selected applications in speech recognition,''
%      \textit{Proceedings of the IEEE}, vol.~77, no.~2, pp.~257-286, 1989.
%    \bibitem[3]{Hastie09-TEO}
%      T.\ Hastie, R.\ Tibshirani, and J.\ Friedman,
%      \textit{The Elements of Statistical Learning -- Data Mining, Inference, and Prediction}.
%      New York: Springer, 2009.
%    \bibitem[4]{YourName16-XXX}
%      F.\ Lastname1, F.\ Lastname2, and F.\ Lastname3,
%      ``Title of your INTERSPEECH 2016 publication,''
%      in \textit{Interspeech 2016 -- 16\textsuperscript{th} Annual Conference of the International Speech Communication Association, September 8–12, San Francisco, California, USA, Proceedings, Proceedings}, 2016, pp.~100--104.
%  \end{thebibliography}

\end{document}